\documentclass[twocolumn,pra,showpacs,amsmath,amssymb]{revtex4}
\usepackage{graphicx}
\usepackage{amssymb,amsthm,amsfonts,amstext}
\usepackage{url}
\def\be{\begin{equation}}
\def\ee{\end{equation}}
\def\bea{\begin{eqnarray}}
\def\eea{\end{eqnarray}}
\def\bma{\begin{mathletters}}
\def\ema{\end{mathletters}}

\def\0{\overline{0}}

\def\q0{\underline{0}}

\def\C{{\mathbb C}}
\def\id{{\mathbb I}}

\def\R{\mathbb{R}}

\def\tr{\mbox{tr}}
\def\one{\leavevmode\hbox{\small1\normalsize\kern-.33em1}}

\def\bra#1{\langle#1|} \def\ket#1{|#1\rangle}

\def\proj#1{\ket{#1}\!\bra{#1}}

\def\id{{\mathbb I}}
\def\tr{\mbox{tr}}

\begin{document}

\title{Bounding the set of finite dimensional quantum correlations}
\author{Miguel Navascu\'es, $^{1}$
        Tam\'as V\'ertesi,$^{2,3}$}
\affiliation{$^1${\it Department of Physics, Bilkent University, Ankara 06800, Turkey}\\
$^2${\it Institute for Nuclear Research, Hungarian Academy of Sciences, H-4001 Debrecen, P.O. Box 51, Hungary}\\
$^3${\it D\'epartement de Physique Th\'eorique, Universit\'e de Gen\`eve, 1211 Gen\`eve, Switzerland}}

\begin{abstract}
We describe a simple method to derive high performance semidefinite programming relaxations for optimizations over complex and real operator algebras in finite dimensional Hilbert spaces. The method is very flexible, easy to program and allows the user to assess the behavior of finite dimensional quantum systems in a number of interesting setups. We use this method to bound the strength of quantum nonlocality in bipartite and tripartite Bell scenarios where the dimension of a subset of the parties is bounded from above. We derive new results in quantum communication complexity and prove the soundness of the prepare-and-measure dimension witnesses introduced in [Phys. Rev. Lett. 105, 230501 (2010)]. Finally, we propose a new dimension witness that can distinguish between classical, real and complex two-level systems.
\end{abstract}

\maketitle
%One of the most astounding predictions of quantum theory is that many physical magnitudes which in classical Physics were regarded as continuous, are actually discretized. Even more surprising is the realization that certain experimental setups can only store a finite amount of information, related to the dimensionality of the underlying quantum system \cite{holevo}.

The realization that certain experimental setups can only store a finite amount of information, related to the dimensionality of the underlying quantum system \cite{holevo}, is one of the most surprising features of quantum theory. Building upon this observation, quantum communication complexity studies the possibility to conduct a distributed computation when we limit the dimension of the quantum systems we allow to exchange \cite{review,review2}. An upper bound on the dimension of the systems transmitted is also the basis of semi-device independent quantum key distribution (QKD) and randomness expansion \cite{semi_qkd, semi_rand}. In the other direction, establishing \emph{lower bounds} on the number of quantum levels that new quantum technologies can effectively control is the key to assess their potential for quantum computation, or as quantum memories. These bounds can be derived from prepare-and measure dimension witnesses \cite{gallego}, such as the ones used in quantum communication complexity and semi-device independent QKD, or even better, through the degree of violation of a Bell inequality \cite{d_wit1,d_wit2,d_wit3,d_wit4}. This last technique has the advantage that classical degrees of freedom have no contribution on the certified dimensionality.

%More recently, the concept of dimensionality has generated a sudden interest, due to the rapid development of modern quantum technologies [REFERENCIAS!!!], whose potential for quantum computation or as a quantum memory depends on the number of quantum levels they can effectively control.

The above motivates the need of characterizing the statistics achievable with quantum systems of a given dimension. In this respect, nowadays we have highly effective variational methods, like see-saw \cite{seesaw1,seesaw}, which allow us to explore the \emph{interior} of the set of $D$-dimensional quantum correlations for fairly high values of $D$. However, we still lack good tools to establish limits on the strength of such correlations, i.e., to characterize this set from the \emph{outside}. So far, all proposed methods to tackle this problem are either very computationally demanding \cite{d_wit1,gonzalo}, cannot be shown to converge \cite{moroder,marcin_SDP} or only apply to particular functionals of the measured probabilities \cite{d_wit2,d_wit3,d_wit4,ours,bowles}.

In this letter, we propose a simple scheme to generate semidefinite programming (SDP) \cite{sdp} relaxations of dimensionally-constrained problems in quantum information theory. Such relaxations, whose working principles stem from noncommutative polynomial optimization theory \cite{siam}, beat all previous methods in almost all conceivable scenarios. For the sake of clarity, we have chosen to illustrate how the scheme works by applying it to specific setups of interest in quantum information theory. A general approach to the noncommutative polynomial optimization problem under dimension constraints, together with an analysis of the convergence of the hierarchies introduced here, will appear elsewhere \cite{future}.

Let us start by considering the problem of introducing dimension constraints in quantum nonlocality. Suppose that we wish to maximize a Bell functional $B(P)=\sum_{x,y}\sum_{a,b}B^{x,y}_{a,b}P(a,b|x,y)$ under the constraint that each party has access to a $D$-level quantum system. That is, we want to solve the problem:

\bea
&&\max \sum_{x,y,a,b}B^{x,y}_{a,b}P(a,b|x,y),\nonumber\\
s.t. && P(a,b|x,y)=\bra{\psi}E^x_a\otimes F^y_b\ket{\psi},
\label{Bell_dim}
\eea

\noindent where $\{E^x_a,F^y_b\}$ are projection operators acting in $\C^D$, with $\sum_aE^x_a=\sum_bF^y_b=\id_D$, and $\ket{\psi}\in \C^{D^2}$.

One way to attack this problem is to simply ignore the dimension restrictions and apply the Navascu\'es-Pironio-Ac\'in (NPA) hierarchy of semidefinite programs for the characterization of quantum correlations \cite{npa, npa2}.

The NPA hierarchy works by reformulating problems such as (\ref{Bell_dim}) as linear optimizations over the underlying system's \emph{moment matrix}. The moment matrix of a quantum system like the one above, with operators $\tilde{E}^x_a=E_a^x\otimes \id,\tilde{F}^y_b=\id\otimes F^y_b$ and state $\ket{\psi}$, is a matrix $\Gamma$ whose rows and columns are labeled by strings of these operators (e.g.: $\id, \tilde{F}^y_b, \tilde{E}^x_a\tilde{E}^{x'}_{a'}$), and such that, for any two strings of operators $u,v$,

\be
\Gamma_{u,v}=\bra{\psi}u^\dagger(\tilde{E}^x_a,\tilde{F}^y_b) v(\tilde{E}^x_a,\tilde{F}^y_b)\ket{\psi}.
\ee

\noindent It can be easily shown that any finite principal submatrix of the moment matrix must be positive semidefinite \cite{npa,npa2}. The operating principle behind the NPA hierarchy is to approximate the set of feasible moment matrices ${\cal R}$ by the set of normalized positive semidefinite matrices which belong to the span of ${\cal R}$. That last condition is enforced by imposing that

\be
\Gamma=\sum_{u}c_uN_u+c^*_uN_{u^\dagger},
\label{span_inf}
\ee

\noindent where $N_u$ is a matrix defined by

\bea
(N_u)_{v,w}=&&1, \mbox{ if } v^\dagger w=u;\nonumber\\
&&0, \mbox{ otherwise}.
\eea

Calling ${\cal S}_{\infty}$ the matrix subspace defined by (\ref{span_inf}), the NPA relaxation to problem (\ref{Bell_dim}) is:

\bea
&&\max \sum_{x,y,a,b}B^{x,y}_{a,b}\Gamma_{\tilde{E}^{x}_a,\tilde{F}^{y}_b}\nonumber\\
s.t. && \Gamma_{\id,\id}=1, \Gamma\geq 0,\Gamma\in {\cal S}_{\infty}.
\label{Bell_NPA}
\eea

\noindent By imposing the positivity condition just over the finite dimensional matrix $\{\Gamma_{u,v}:\mbox{length}(u),\mbox{length}(v)\leq n\}$, we end up with a finite dimensional problem, the \emph{$n^{th}$-order relaxation} of (\ref{Bell_NPA}). This happens to be a semidefinite program (SDP), a class of optimization problems for which plenty of efficient numerical tools are available \cite{sdp}.

Although a sound relaxation of problem (\ref{Bell_dim}), the NPA hierarchy is not sensitive to the dimensionality parameter $D$, and so it cannot be used to derive dimension witnesses. The key to go beyond the NPA approximation is to acknowledge that ${\cal S}_{\infty}$ does not capture all linear restrictions present in moment matrices arising from systems of dimension $D$. We will incorporate dimension constraints to the SDP problem (\ref{Bell_NPA}) by characterizing exactly the span of such a set of matrices, or, more precisely, its $n^{th}$-order truncation. That is, we will identify a \emph{minimal} basis of matrices $\{M_j\}_{j=1}^N$ such that any truncated feasible moment matrix $\Gamma$ of order $n$ arising from a $D$-dimensional quantum system can be expressed as $\Gamma=\sum_{j=1}^N c_j M_j$.

For convenience, we start by making an assumption on the \emph{ranks} of the optimal projectors $\{E^x_a,F^y_b\}$. Calling ${\cal S}_{D,\vec{r}}$ the set of all feasible moment matrices with $\mbox{rank}(E^x_a)=r^{A,x}_a,\mbox{rank}(F^y_b)=r^{B,y}_b$, the problem we wish to solve is:

\bea
&&\max \sum_{x,y,a,b}B^{x,y}_{a,b}\Gamma_{\tilde{E}^{x}_a,\tilde{F}^{y}_b}\nonumber\\
s.t. && \Gamma_{\id,\id}=1, \Gamma\geq 0,\Gamma\in{\cal S}_{D,\vec{r}}.
\label{Bell_dim_rel}
\eea

\noindent In order to conduct its $n^{th}$-order relaxation, we must determine the projection ${\cal S}_{D,\vec{r}}^n$ of ${\cal S}_{D,\vec{r}}$ onto the space of $n^{th}$-order moment matrices.

To that end, we generate randomly quantum states $\proj{\psi^{j}}\in B(\C^{D^2})$ and projection operators $\{E^{x,j}_a,F^{y,j}_b\}\subset B(\C^D)$, with $\mbox{rank}(E^{x,j}_a)=r_{x,a}^A,\mbox{rank}(F^{y,j}_b)=r^{B}_{y,b}$. For each tuple $(\proj{\psi^{j}},E^{x,j}_a,F^{y,j}_b)$ of feasible state and projectors, we build the corresponding $n^{th}$ order moment matrix

\be
\Gamma^{j}_{u,w}=\bra{\psi^{j}}u(\tilde{E}^{x,j}_a,\tilde{F}^{y,j}_b)^\dagger w(\tilde{E}^{x,j}_a,\tilde{F}^{y,j}_b)\ket{\psi^{j}},
\ee

\noindent where $\tilde{E}^{x,j}_a=E^{x,j}_a\otimes \id_D$, $\tilde{F}^{y,j}_b=\id_D\otimes F^{y,j}_b$ and $u,w$ range over all strings of length smaller than or equal to $n$. We thus get a sequence of random feasible moment matrices $\Gamma^{1}, \Gamma^{2},...$. Since on one hand we are only interested in linear combinations of real entries of the system moment matrix (namely, $\{\Gamma_{\tilde{E}^x_a,\tilde{F}^y_b}\}$) and, on the other hand, given a feasible tuple $(\psi^{j},E^{x,j}_a,F^{y,j}_b)$, its complex conjugate $(\psi^{j},E^{x,j}_a,F^{y,j}_b)^*$ is also feasible, it is enough to consider the real part of the above sequence, i.e., $\mbox{Re}(\Gamma^{1}), \mbox{Re}(\Gamma^{2}),...$.

Adopting the Hilbert-Schmidt scalar product $\langle A,B\rangle=\tr(A^\dagger B)$, one can apply the Gram-Schmidt process to this sequence of real moment matrices in order to obtain an orthogonal basis $\tilde{M}_1,\tilde{M}_2,...$ for the space spanned by such matrices. We will notice that, for some number $N$, $\tilde{M}_{N+1}=0$, up to numerical precision. This is the point to terminate the Gram-Schmidt process and define the normalized matrices $\{M_j\equiv \frac{\tilde{M}_j}{\sqrt{\tr{\tilde{M}_j^2}}}:j=1,...,N\}$. It is easy to see that, even though the matrix basis $\{M_j\}_{j=1}^N$ was obtained randomly, the space it represents is always the same, namely, the intersection of ${\cal S}_{D,\vec{r}}^n$ with the set of real symmetric matrices\footnote{Note, however, that, due to rounding errors during the execution of the numerical calculations, the matrix basis generated by the computer is just an approximation to ${\cal S}_{D,\vec{r}}^n$. During the course of this work we observed that this effect sometimes led to `upper bounds' \emph{smaller} than the best known lower bound by an amount of order $10^{-7}$. These numerical paradoxes, also reported in other SDP hierarchies \cite{perturb1,perturb2}, go away by increasing the computer precision.}. %\footnote{Indeed, let $N=\mbox{dim}({\cal S}_{D,\vec{r}}^n)$, and suppose that $\tilde{M}^1,...,\tilde{M}^{j-1}$ are non-zero, with $j\leq N$. If we generate quantum states and projectors in each instance $j$ by sampling a vector $\vec{z}_j$ of random and independent complex entries over which we apply Gram-Schmidt (to ensure orthogonality of the projector eigenvectors), then the entries of the matrix $\tilde{M}^{j}$ will be rational functions of $\vec{z}_j,(\vec{z}_j)^*$. Moreover, since $j\leq N$, there exists a choice of $\vec{z}^j$ such that $\tilde{M}_j(\vec{z}^j)\not=0$. Since the probability that a non-zero rational function vanishes when evaluated randomly is zero, we conclude that $\tilde{M}_j$ will always be non-zero. On the other hand, $\tilde{M}_1=\mbox{Re}(\Gamma^j)\not=0$, so by induction we have that $N$ randomly chosen moment matrices will span ${\cal S}_{D,\vec{r}}^n$ with certainty. Consequently, $\tilde{M}_{N+1}=0$, indicating when to stop the procedure.}.

A brief note on strict feasibility is in order. One can show that, even if we eliminate one projection operator from each measurement as in \cite{npa,npa2} to remove operator dependencies, for high enough relaxation orders $n$ there are no strictly feasible points for problem (\ref{Bell_dim_rel}) \cite{future}. In other words: there is no positive definite $s\times s$ matrix $\Gamma$ satisfying the above constraints. This poses a problem for the implementation of (\ref{Bell_dim_rel}), since many SDP solvers require strict feasiblility to operate. An easy way to circumvent this issue is to compute the matrix $G\equiv\frac{1}{N}\sum_{i=1}^N Re(\Gamma^{(i)})$ and find an isometry $V:\mbox{supp}(G)\to\R^{s}$. Then, the positive semidefinite condition over $\Gamma$ can be replaced by $V\Gamma V^\dagger\geq 0$, which, by construction, admits a strictly feasible point.

We have just described how to perform in practice the $n^{th}$ order SDP relaxation of problem (\ref{Bell_dim_rel}), which, in turn, is a relaxation of a rank-constrained version of problem (\ref{Bell_dim}). Taking the maximum over all possible rank combinations $\vec{r}$, we obtain an upper bound on the solution of (\ref{Bell_dim}).

At this point, the reader will probably wonder whether this method is actually useful for the kind of problems we usually encounter in quantum information theory. Hence, we conducted a number of optimizations over the set of $D$-dimensional quantum correlations in order to assess its performance. Such numerical computations, as well as all subsequent ones presented in this paper, were carried out with the MATLAB packages YALMIP \cite{yalmip} and the SDP solvers SeDuMi \cite{sedumi} and Mosek \cite{mosek}.

First we considered the $I_{3322}$ inequality, a three-setting bipartite two-outcome Bell inequality from the $I_{NN22}$ family, defined in \cite{I3322}. Recently, it has been proven that qubit systems are not enough to attain the quantum maximum $\sim0.2509$ \cite{gonzalo,moroder}. Rather, the best value in $\C^2\times\C^2$ systems is 0.25. Using (\ref{Bell_dim_rel}), we certified up to 7 significant digits that the maximum is 0.25 in dimensions $\C^3\times\C^3$ as well. The computations, which were performed on a normal desktop PC, took about 5 minutes for a fixed rank combination of measurements.

We then switched to the four-setting Bell inequality defined in \cite{gonzalo} by Eq.~(19). 
Its maximal violation in $\C^2\times\C^2$ systems has been upper bounded by 5.8515 \cite{gonzalo}. However, this upper bound turns out to be not tight: using our new tool one can certify a value of 5.8310, which can be matched by see-saw variational methods. By raising the dimension to $\C^3\times\C^3$, one obtains the same number again, which must be compared to the maximum value of $5.9907$, achievable in $\C^4\times\C^4$ systems.

\vspace{10pt}

Note that relaxation (\ref{Bell_dim_rel}) is only valid under the assumption that both parties are conducting projective measurements. More general measurements are modeled in quantum theory via Positive Operator Valued Measures (POVMs), i.e., by a collection of operators $\{M^x_a\}\subset B(\C^D)$ with $M^x_a\geq 0$, $\sum_aM^x_a=\id_D$. It so happens that, for two-outcome measurements, the extreme points of the POVM set are given by projective operators. Relaxation (\ref{Bell_dim_rel}) is hence sound in all such scenarios.

In order to study more complex measurement setups, we can exploit the fact that any $d$-outcome POVM in dimension $D$ can be viewed as a projective measurement in an extended Hilbert space $\C^d\otimes \C^D$ \cite{nielsen}. In this dilation picture, Alice and Bob's state would be of the form $\proj{0}_{A'}\otimes\proj{0}_{B'}\otimes\proj{\psi}_{AB}$, with $\mbox{dim}(A')=\mbox{dim}(B')=d$, $\mbox{dim}(A)=\mbox{dim}(B)=D$, while Alice's local measurement projectors are given by $E^x_a= U^x(\proj{a}\otimes \id_D)(U^x)^\dagger$, where $U^x\in B(A'\otimes A)$ is an arbitrary unitary operator (and similarly for Bob). A feasible moment matrix for this system would thus be generated by sampling random unitaries $U_x,V_y$ and states $\Psi$. In this scheme it may be convenient to introduce two different `identity operators' in our moment matrices. One of them would be the genuine identity $\id_{A'B'}\otimes \id_{AB}$ on Alice and Bob's target and ancillary states. The other one would be the projector $\proj{0}^{\otimes 2}_{A'B'}\otimes \id_{AB}$ onto Alice and Bob's target space $AB$.

To test the efficiency of the above method, we picked the Pironio-Bell inequality \cite{d_wit1,pironio}, which is the simplest tight Bell inequality beyond two-outcome inequalities. Here Alice has three binary-outcome measurement settings, whereas Bob's first setting has binary outcomes and his second setting has ternary outcomes. By allowing general POVM measurements on Bob's second setting, we recover the two-qubit quantum maximum $(\sqrt 2 - 1)/2\simeq0.2071$ up to 8-digit precision on level 3 of the hierarchy. Note the overall quantum maximum is a larger value of 0.2532 which can be attained with two-qutrit systems \cite{d_wit1}.

The method sketched above can be easily extended to deal with multipartite Bell scenarios where only a subset of the parties has limited dimensionality. Consider, for instance, a tripartite scenario where the dimensionality of Charlie's system is $D$, but Alice and Bob's measurement devices are otherwise unconstrained. We want to generate a basis for the corresponding space of truncated moment matrices, with rows and columns labeled by strings of operators of the form $u(AB)v(C)$, where $u(AB)$ ($v(C)$) denotes a string of Alice and Bob's (Charlie's) operators of length at most $n_{AB}$ ($n_C$).

The key is to realize that, in a multipartite (complex) Hilbert space, the space of feasible moment matrices is spanned by moment matrices corresponding to separable states. Hence, in order to attack this problem, we start by generating a sequence of \emph{complex} $D$-dimensional moment matrices for Charlie's system alone. After applying Gram-Schmidt to these complex matrices, we obtain the basis of Hermitian matrices $\{M_j\}_{j=1}^N$. Next, we generate a basis for Alice and Bob's moment matrices. Since their dimension is unconstrained, we invoke eq. (\ref{span_inf}). The overall moment matrix for the whole system can then be expressed as $\Gamma=\sum_{u,j} (c_{u,j}N_u+c_{u,j}^*N_{u^\dagger})\otimes M_j$.

If, as before, we are just interested in optimizing a real linear combination of real entries of $\Gamma$ -corresponding to the measured probabilities $P(a,b,c|x,y,z)$-, we can take the real part of the above matrix. We hence arrive at an SDP involving real matrices and variables.

To check its performance, we chose the following tripartite Bell inequality:
\be
CHSH_{AC}+CHSH_{BC'}\le 4,
\label{quasi_toner}
\ee
where $CHSH_{AC}=A_1C_1+A_1C_2+A_2C_1-A_2C_2$ is the famous CHSH expression~\cite{chsh}. Eq. (\ref{quasi_toner}) is similar to the Bell inequality studied in Ref.~\cite{TV06}. Note, however, that here Charlie's measurement settings $C$ and $C'$ are different (i.e., he has four settings). Hence, in principle, we cannot restrict to states where Charlie has support on a qubit in order to compute the maximal quantum violation of (\ref{quasi_toner}). Actually, by running an SDP for the case where Charlie holds a qubit, we find the (obviously tight) upper bound of $2+2\sqrt 2$ up to 8-digit precision. This value shall be contrasted with the overall quantum maximum of $4\sqrt 2$, attainable in four-level quantum systems.

Limiting quantum nonlocality under dimension constraints is not the only interesting problem in quantum information that can be solved with the above scheme. Consider, for instance, the problem of bounding the efficiency of quantum communication complexity.

Two parties, call them Alice and Bob, receive the inputs $x,y$, respectively. They wish to compute the Boolean function $f(x,y)\in \{0,1\}$, for which purpose Alice is allowed to transmit Bob a $D$-level quantum system. The question is: given a prior distribution of the inputs $p(x,y)$, what is the maximum probability that Bob guesses the value of $f(x,y)$?

This scenario can be modeled by assuming that Bob performs a binary measurement $F^y_b$ over the state $\rho_x$ sent by Alice. The outcome $b\in \{0,1\}$ will be Bob's guess, which he will output with probability $P(b|x,y)=\tr(\rho_xF^y_b)$. The problem we wish to solve is hence:

\bea
\max &&\sum_{x,y}p(x,y) \tr(\rho_xF^y_{f(x,y)}),\nonumber\\
s.t. &&\tr(\rho_x)=1,\rho_x^2=\rho_x, (F^y_b)^2=F^y_b,\nonumber\\
&&\rho_x, F^y_b\in B(\C^D).
\label{q_comm}
\eea

\noindent Here we have exploited the fact that the extreme points of the distributions $P(b|x,y)$ are generated by pure states and projective measurements. Note that the maximal value of prepare-and-measure dimension witnesses, as defined in \cite{gallego}, can also be expressed as a linear optimization over the set of feasible probabilities $P(b|x,y)$.

There are many ways to reformulate problem (\ref{q_comm}), e.g.: by modeling the preparation device via measurements on one side of a maximally entangled state, as in \cite{marcin_SDP}. Each of them leads to a different hierarchy of SDP relaxations. Here we study the most obvious choice: namely, we regard our reference state as the unnormalized maximally mixed state in dimension $D$; and $\rho_x$, as rank-1 projectors. Hence we obtain our random basis by choosing randomly the projectors $\rho^j_x,F^{y,j}_b\in B(\C^D)$, with $\mbox{rank}(\rho^j_x)=1, \mbox{rank}(F^{y,j}_b)=r^y_b$ and using them to construct the moment matrices $\Gamma^j_{u,v}=\tr\{u(\rho^j_x,F^{y,j}_b)^\dagger v(\rho^j_x,F^{y,j}_b)\}$. Denoting by ${\cal T}_{D,\vec{r}}$ the span of the real part of all such matrices, the resulting program is:

\bea
\max &&\sum_{x,y}P(x,y)\Gamma_{\rho_x,F^y_{f(x,y)}}\nonumber\\
s.t. &&\Gamma_{\id,\id}=D,\Gamma\geq 0,\Gamma\in {\cal T}_{D,\vec{r}}.
\label{tracial}
\eea

Let us explore how relaxations of the problem above perform in practice. In a Quantum Random Access Code (QRAC) \cite{QRAC}, the inputs $\vec{x},y$ can take values in $\{0,1\}^k$ and $\{1,...,k\}$, respectively, and the function to compute is $f(\vec{x},y)=x_y$. If the inputs are distributed independently and uniformly and Alice is allowed to transmit Bob a $D$-level quantum system, the average success probability of the optimal \emph{$k\to\log_2(D)$ QRAC} is usually denoted as $P_{\max}(k\to \log_2(D))$ \cite{marcin_QRAC}.

It was previously known that $P_{\max}(2\to 1)=1/2+\sqrt{2}/4$ \cite{QRAC}. Actually, this is the value given by program (\ref{tracial}) at order $n=2$, up to computer precision. Likewise, for $D=3$, i.e., when we allow Alice to transmit a qutrit, program (\ref{tracial}) at the same order gives $P_{\max}\leq 0.90450850$, which is equal up to numerical precision to the lower bound obtained via see-saw methods.

The second-order relaxation of (\ref{tracial}) also performs well when we increase $D$ and $k$. Table \ref{QRAC_3} shows bounds on the average success probability for QRAC $3\to \log_2(D)$ for different values of $D$, computed via program (\ref{tracial}) in a normal desktop (using the solver Mosek \cite{mosek}). It is worth noting that, except for the cases $D=5,6$, with gaps between the upper (UB) and lower bounds (LB) of the order of $10^{-6}$ and $10^{-3}$, respectively, the values obtained via see-saw and (\ref{tracial}) are equal up to numerical precision.

\begin{table}
\label{QRAC_3}
\begin{center}\begin{tabular}{|c|c|c|c|c|c|c|c|}
\hline
D& 2& 3& 4& 5& 6& 7\\
\hline
LB& 0.788675& 0.832273& 0.908248& 0.924431& 0.951184& 0.969841\\
UB& 0.788675& 0.832273& 0.908248& 0.924445& 0.954123& 0.969841\\

  \hline
\end{tabular}
\caption{Lower and upper bounds on $P_{\max}(3\to\log_2(D))$.}
\end{center}
\end{table}

We also used program (\ref{tracial}) to re-compute the optimal quantum value of the dimension witnesses $I_N$ defined in [\cite{gallego}, table I]. We found that the second relaxation produced upper bounds on the maximal violation of $I_N$ that matched the lower bounds obtained via see-saw for $N=3,4$ and $D=2,3$. To appreciate the importance of these calculations, note that the conclusions of the experimental paper \cite{gallego_exp} relied on the conjecture that the lower bounds for $I_4$ provided in \cite{gallego} were optimal.

Finally, we tested the behavior of program (\ref{tracial}) to bound the set of accessible probabilities in scenarios where measurements have more than two outcomes. Arbitrarily, we chose the reference state of our moment matrix to be $\id_d\otimes \id_D$, with $d$ being the number of outcomes (note that we could have chosen $\proj{0}\otimes\id_D$ as well), and renormalized our moment matrix accordingly.

Now, suppose that Alice and Bob wish to compute the function $f(x,y)=(1+2\delta_{x,3})y-x \mbox{ (mod 3)}$, with $x=0,1,2,3$ and $y=0,1$. By dilating the three-outcome measurements $\{F^y_a:a=0,1,2\}$ to three rank-2 projectors in $B(\C^3\otimes\C^2)$, both (the renormalized version of) program (\ref{tracial}) at $k=2$ and see-saw give a maximum probability of success of $3/4$ when $D=2$, the same as the classical bit value. For $D=3$, program (\ref{tracial}) returns $0.904508$, again coincident with the see-saw value.

%With such a model, it can be proven \cite{future} that the presence of the `extra identity' $\proj{0}\otimes \id_D$ in the moment matrix ensures that the second-order relaxation achieves the exact value of the dimension witnesses proposed in \cite{ours} for scenarios with $N$ preparations and a single $N$-outcome measurement. We verified that removing this operator from the moment matrix results in an upper bound of $\approx 0.7888$ for the $N=3, D=2$ witness, well above the exact solution of $2/3$ \cite{ours}.

So far we have been interested in bounding the behavior of complex quantum mechanical systems, but nothing prevents us from applying the same ideas to characterize the properties of \emph{real} quantum mechanical systems as well. Consider the dimension witness $V_4$, defined in \cite{ours}, and take $D=2$. Running the SDP for the case of complex qubits, we recover the upper bound $Q_{2C}=2\sqrt 6$. This bound is tight and can be saturated via SIC POVM's \cite{ours}. In the real qubit case, though, we obtain the upper bound $Q_{2R}=2(\sqrt 2 + 1)$, also tight. We obtained both results in a few seconds on a normal desktop PC by using a relaxation level intermediate between two and three.

\vspace{10pt}

\noindent\emph{Conclusion}

\vspace{5pt}

We have described a simple method to derive SDP relaxations for optimizations over operator algebras under dimension constraints. This method allows us to attack a number of relevant problems in quantum information theory, such as the characterization of quantum nonlocality under dimension constraints or the determination of the quantum communication complexity of arbitrary Boolean functions. As we saw, the method even distinguishes between real and complex algebras, and hence it can be used to certify that a given experimental setup has control over a complex $D$-dimensional space.

%In a forthcoming paper \cite{future}, we will present a general approach to noncommutative polynomial optimization theory under dimension constraints, where we will show how to model arbitrary operator \emph{inequalities} and prove the completeness of some of the semidefinite programming hierarchies sketched here.

Note that one can also use the non-deterministic algorithms sketched above to identify the space spanned by \emph{tensor products} $\Gamma^{\otimes n}$ of $n$ moment matrices. By imposing the existence of a symmetric separable decomposition (rather than just positive semidefiniteness) over all matrices belonging to such a space, we hence obtain a non-trivial relaxation for the convex hull of \emph{$n$-degree polynomials} of the system's average values. It would be interesting to explore whether this scheme leads to good outer approximations of the (non-convex) set of $D$-dimensional quantum correlations.

\vspace{10pt}
\noindent\emph{Acknowledgements}

We thank A. Winter for useful discussions. M.N. acknowledges the European Commission (EC) STREP "RAQUEL", as well as the MINECO project FIS2008-01236, with the support of FEDER funds. T.V. acknowledges financial support from a J\'anos Bolyai Grant of the Hungarian Academy of Sciences, the Hungarian National Research Fund OTKA (K111734), and SEFRI (COST action MP1006).

\end{document}